\documentclass[12pt,preprint]{aastex}

\shorttitle{Ultra-bright SMG in the SMC}
\shortauthors{Takekoshi et al.}

\begin{document}


\title{Detection of an Ultra-bright Submillimeter Galaxy behind the Small Magellanic Cloud\thanks{{\it Herschel} is an ESA space observatory with science instruments provided by European-led Principal Investigator consortia and with important participation from NASA.}}


\author{Tatsuya Takekoshi\altaffilmark{1,2}, Yoichi Tamura\altaffilmark{3}, Tetsuhiro~Minamidani\altaffilmark{1,2,3,4}, Kotaro~Kohno\altaffilmark{3,5}, Taira~Oogi\altaffilmark{4}, Kazuo~Sorai\altaffilmark{1,4}, Asao~Habe\altaffilmark{1,4}, Hajime~Ezawa\altaffilmark{6}, Tai~Oshima\altaffilmark{2}, Kimberly~S.~Scott\altaffilmark{7}, Jason~E.~Austermann\altaffilmark{8}, Shinya~Komugi\altaffilmark{6,9}, Tomoka~Tosaki\altaffilmark{10}, Norikazu~Mizuno\altaffilmark{6}, Erik~Muller\altaffilmark{6}, Akiko~Kawamura\altaffilmark{6}, Toshikazu~Onishi\altaffilmark{11}, Yasuo~Fukui\altaffilmark{12}, Hiroshi~Matsuo\altaffilmark{13}, Itziar~Aretxaga\altaffilmark{14}, David~H.~Hughes\altaffilmark{14}, Ryohei~Kawabe\altaffilmark{2,6,9}, Grant~W.~Wilson\altaffilmark{15}, and Min~S.~Yun\altaffilmark{15}}






\altaffiltext{1}{Department of Cosmosciences, Graduate School of Science, Hokkaido University, Sapporo 060-0810, Japan}
\altaffiltext{2}{Nobeyama Radio Observatory, National Astronomical Observatory of Japan, Minamimaki, Minamisaku, Nagano 384-1305, Japan}
\altaffiltext{3}{Institute of Astronomy, The University of Tokyo, Osawa, Mitaka, Tokyo 181-0015, Japan}
\altaffiltext{4}{Department of Physics, Faculty of Science, Hokkaido University, Sapporo 060-0810, Japan}
\altaffiltext{5}{Research Center for the Early Universe, University of Tokyo, 7-3-1 Hongo, Bunkyo, Tokyo 113-0033, Japan}
\altaffiltext{6}{Chile Observatory, National Astronomical Observatory, Mitaka, Tokyo 181-8588, Japan}
\altaffiltext{7}{North American ALMA Science Center, National Radio Astronomy Observatory, Charlottesville, VA 22903, USA}
\altaffiltext{8}{Center for Astrophysics and Space Astronomy, University of Colorado, Boulder, CO 80309, USA}
\altaffiltext{9}{Joint ALMA Observatory, Vitacura, Santiago, Chile}
\altaffiltext{10}{Joetsu University of Education, Joetsu, Niigata 943-8512, Japan}
\altaffiltext{11}{Department of Physical Science, Osaka Prefecture University, Gakuen 1-1, Sakai, 599-8531 Osaka, Japan}
\altaffiltext{12}{Department of Astrophysics, Nagoya University, Chikusa-ku, Nagoya 464-8602, Japan}
\altaffiltext{13}{Advanced Technology Center, National Astronomical Observatory of Japan, Mitaka, Tokyo 181-8588, Japan}
\altaffiltext{14}{Instituto Nacional de Astrof\'{i}sica, \'{O}ptica y Electr\'{o}nica (INAOE), 72000 Puebla, Mexico}
\altaffiltext{15}{Department of Astronomy, University of Massachusetts, Amherst, MA 01003, USA}


\begin{abstract}
We report the discovery of a new ultra-bright submillimeter galaxy (SMG) behind the Small Magellanic Cloud (SMC).
This SMG is detected as a $43.3{\pm}8.4$~mJy point source (MM~J01071$-$7302, hereafter MMJ0107) in the 1.1-mm continuum survey of the SMC by AzTEC on the ASTE telescope.
MMJ0107 is also detected in the radio (843~MHz), {\it Herschel}/SPIRE, {\it Spitzer} MIPS~24 $\mathrm{{\mu}m}$, all IRAC bands, {\it Wide-field Infrared Survey Explorer}, and near-infrared ($J$, $H$, $K_S$).
We find an optical ($U$, $B$, $V$) source, which might be the lensing object, at the distance of $1.4''$ from near-infrared and IRAC sources.
Photometric redshift estimates for the SMG using representative spectral energy distribution templates show the redshifts of 1.4--3.9.
We estimate total far-infrared luminosity of $(0.3$--$2.2) \times 10^{14}~\mu^{-1}~L_\sun$ and a star formation rate of $5600$--$39,000~\mu^{-1}~{M_\sun~yr^{-1}}$, where $\mu$ is the gravitational magnification factor.
This apparent extreme star formation activity is likely explained by a highly magnified gravitational lens system.
\end{abstract}


\keywords{galaxies: formation --- galaxies: high-redshift --- galaxies: starburst --- infrared: galaxies --- Magellanic Clouds}misumi



\section{Introduction}
Advances in millimeter (mm) and submillimeter (submm) continuum observations provide a chance to investigate the extreme star formation activity in the early Universe via observations of submm-bright galaxies \citep[hereafter SMGs; e.g.,][]{2002PhR...369..111B}.
In SMGs, dust grains that are warmed by the formation of young massive stars (total star formation rate (SFR) $\sim$ 100 -- 1000 ${M_\sun~\mathrm{yr^{-1}}}$) emit strong far-infrared (FIR) thermal radiation ($L_\mathrm{FIR} \sim$ $10^{12}$ -- $10^{13}~{L_\sun}$), which is redshifted into the submm band at high redshift.

Strong gravitational lensing magnification by foreground galaxies aids detection of the distant star-forming galaxies.
Recent wide field mm/submm surveys have discovered the ultra-bright population of SMGs.
An early result of {\it Herschel}-ATLAS reports the detection of five ultra-bright SMGs ($S_{500\mathrm{{\mu}m}}>100$~mJy) over 14.4 $\deg^2$, and  the source counts of $S_{500\mathrm{{\mu}m}}>100$~mJy are explained by the strongly lensed SMGs \citep{2010Sci...330..800N}.
The {\it Herschel} Multi-tiered Extragalactic Survey also found 13 candidates of strongly lensed SMGs ($S_{500\mathrm{{\mu}m}} > 100$~mJy) from the 95 $\deg^2$ data \citep{2013ApJ...762...59W}.
The objects found by {\it Herschel} were selected at the submm and FIR wavelengths shorter than 500 $\micron$, so cannot be used to constrain the overall redshift distribution of the number densities.
Thus, it is also important to select strongly lensed SMGs at longer wavelengths in the (sub)mm to place a constraint on the high-redshift number densities.
The South Pole Telescope (SPT) observations at 1.4 and 2.0~mm provide 20 dust-dominated sources that are brighter than 10~mJy at 1.4~mm in the 87 $\deg^2$ field \citep{2010ApJ...719..763V}.
APEX LABOCA (870 $\micron$) and SABOCA (350  $\micron$) follow-up observations provide 11 dust-dominated sources found from the 200 $\deg^2$ SPT survey constrain a median redshift of ${\bar z_\mathrm{photo}}=3.0$ \citep{2012ApJ...756..101G}.
ALMA observations toward 26 samples in the 1300 $\deg^2$ data determine the spectroscopic redshifts of 23 sources \citep[${\bar z_\mathrm{spec}}=3.6$;][]{2013Natur.495..344V}.
These results reveal that SPT-selected dusty starburst galaxies are a more distant population of strongly lensed SMGs than {\it Herschel}-selected samples \citep[${\bar z_\mathrm{photo}}=2.0$ and ${\bar z_\mathrm{spec}}=2.6$;][]{2010Sci...330..800N}.

It is noteworthy that 1.1 mm data can also select more distant SMGs than {\it Herschel}, and further, that they suffer less from contamination of unlensed synchrotron sources than SPT surveys.
Whereas AzTEC \citep{2008MNRAS.386..807W} 1.1 mm observations have put a strong constraint on the number counts for relatively normal SMGs \citep[e.g.][]{2011MNRAS.411..102H, 2012MNRAS.423..575S}, no $S_{1.1\mathrm{mm}} >$ 20~mJy SMG has been identified in blank-field surveys except for Orochi \citep[$S_{1.1\mathrm{mm}}$=37~mJy;][]{2011MNRAS.415.3081I}.
To constrain the bright end of the 1.1 mm counts, it is important to search for ultra-bright SMGs from 1.1 mm data to reveal the actual number counts in all flux ranges.

In this Letter, we report the discovery of a new ultra-bright SMG behind the Small Magellanic Cloud (SMC) based on the 1.1 mm observation toward the SMC Wing region ($1.21~\deg^2$) using AzTEC on the ASTE telescope (T. Takekoshi et al., in preparation).
We also investigate the star formation activity and the lensing properties of the discovered SMG using archival data at various wavelengths.
Throughout the Letter, we adopt cosmological parameters of $\Omega_M=0.3$, $\Omega_\Lambda=0.7$, and $H_0=70~\mathrm{km~s^{-1}~Mpc^{-1}}$, and the distance to the SMC of 60~kpc \citep[e.g.,][]{2005MNRAS.357..304H}.

\section{The data} 
Continuum observations at 1.1~mm toward the SMC were performed with the AzTEC camera \citep{2008MNRAS.386..807W} mounted on the ASTE telescope \citep{2004SPIE.5489..763E, ezawa2008new} in Atacama, Chile, from 2008 October 7 to December 26. 
AzTEC is a 144 pixel bolometer camera with a bandwidth of 49~GHz and a center frequency of 270~GHz.
The angular resolution is $28''$ in full width at half-maxima (FWHM). 
We covered a 4.5 $\deg^2$ region of the SMC by connecting four $1.1^\circ \times1.1^\circ$ (position angle of $20^\circ$) patches of raster scans.
One of the patches, centered at $\mathrm{\alpha}(J2000)=01^\mathrm{h}13^\mathrm{m}00^\mathrm{s}$ and $\mathrm{\delta}(J2000)=-73^\circ00'00''$, is less contaminated by dust emission from the SMC than the other three.
This allows us to unambiguously search for extragalactic objects.
Pointing observations with quasar J2355$-$534 were performed every two hours across the observations.
The resulting pointing accuracy applied a pointing model is better than $3''$ at 1$\sigma$ confidence level \citep{2008MNRAS.390.1061W}.
Uranus was observed once every night for determining the absolute calibration and the point-spread functions of each bolometer pixel. 
Flux calibration error has an uncertainty of 8\% \citep{2008MNRAS.390.1061W}. 
The noise level is $\sim$7~$\mathrm{mJy~beam^{-1}}$, which is sufficient to detect ultra-bright SMGs ($S_{1.1\mathrm{mm}}>$ 35~mJy) over 5$\sigma$.
The correlated noise removal is made with a principle component analysis technique \citep{2005ApJ...623..742L,2008MNRAS.385.2225S}, which is optimal for point source analysis.
All of the AzTEC data were reduced using the standard customized data reduction pipeline written in IDL, which is described in \citet{2008MNRAS.385.2225S} and \citet{2012MNRAS.423..529D}.
The point response function after the analysis gives an FWHM of $35''$.

We have access to multi-wavelength data obtained by observations from radio to optical wavelengths.
Radio continuum data at 843~MHz \citep{1998PASA...15..280T} taken with the Molonglo Observatory Synthesis Telescope (MOST) and at 1.4, 2.8, 4.8, and 8.6~GHz \citep{2002MNRAS.335.1085F, 2010AJ....140.1511D} with the Australia Telescope Compact Array (ATCA) are available.
\citet{2011SerAJ.183..103W} has published a radio source catalog at 0.8, 1.4, and 2.8~GHz using these data.
We also use the {\it Herschel}/SPIRE \citep{2010A&A...518L...3G} 250, 350, and 500 $\mathrm{{\mu}m}$ images taken in the HERITAGE project \citep[][; observation ID 1342205055, observed on 2010 September 24]{2010A&A...518L..71M}.
The catalog and images of {\it Spitzer} MIPS (24, 70, and 160 $\mathrm{{\mu}m}$) and IRAC (3.6, 4.5, 5.8, and 8 $\mathrm{{\mu}m}$) are provided by the SAGE-SMC program \citep{2011AJ....142..102G}.
The {\it Wide-field Infrared Survey Explorer} \citep[{\it WISE};][]{2010AJ....140.1868W} has 3.4, 4.6, 12, and 22 $\mathrm{{\mu}m}$ data for the all sky.
Near-IR point source catalog and images of $J$, $H$, and $K_S$ bands were obtained by the SIRIUS camera on the IRSF 1.4 m telescope \citep{2007PASJ...59..615K}.
The optical photometric catalog of $U$, $B$, $V$, and $I$ bands is available from \citet{2002AJ....123..855Z}, which is obtained by the Las Campanas Swope telescope and the Great Circle Camera. The spatial resolutions of each observation are summarized in Table~\ref{table:flux}.

\section{Results}
\label{l:result}
We select candidate ultra-bright SMGs from the 1.1 mm continuum image of the SMC Wing region.
We require $\geq~5\sigma$ ($\sim$35~mJy) for detection, and exclude seven sources that have been identified as massive star-forming regions in the SMC.
Consequently, the 1.1 mm continuum image in the SMC Wing region contains two extragalactic candidates with 1.1 mm flux densities above 35~mJy and no counterparts embedded in the SMC.
We denominate these sources as MM~J01115$-$7302 and MM~J01071$-$7302 (hereafter MMJ0111 and MMJ0107).

We use SIMBAD to search for the counterparts of the two candidates.  
MMJ0111, which is located at $\mathrm{\alpha}(J2000)=01^\mathrm{h}11^\mathrm{m}32_\cdot^\mathrm{s}4$ and $\mathrm{\delta}(J2000)=-73^\circ02'12''$ (1.1 mm flux density of $48.4{\pm}8.2$~mJy), is listed in the 6dF Galaxy Survey catalog \citep[g0111325$-$730210,][]{2009MNRAS.399..683J}, and the redshift of the galaxy is $z=0.06660$.
\citet{2011MNRAS.417.2651M} classify the galaxy as a radio-loud active galactic nucleus (AGN), based on identification of the optical counterpart of the Australia Telescope 20~GHz Survey source.

\begin{figure}
\epsscale{1.00}
\plotone{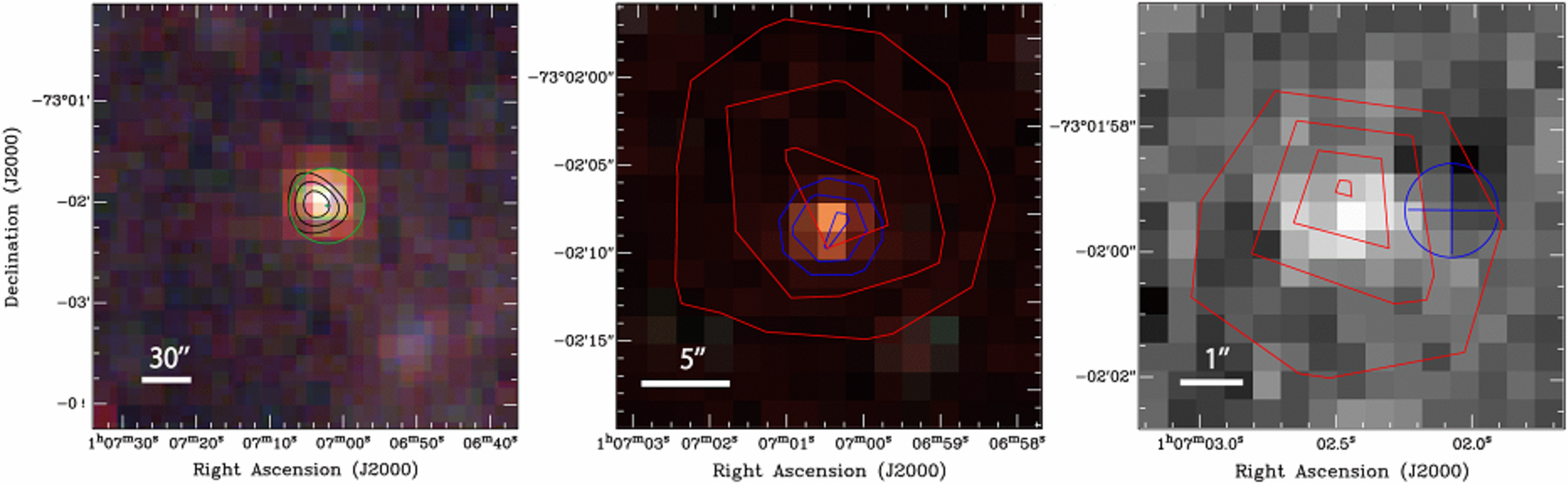}
\caption{Multi-wavelength views of MMJ0107.
{Left}: false color image of {\it Herschel}/SPIRE (500, 350, and 250 $\mathrm{{\mu}m}$ for red, green, and blue, respectively) is shown with contours of AzTEC/ASTE 1.1 mm (black, 3$\sigma$, 4$\sigma$, and 5$\sigma$). The green cross and circle show the position (cross length corresponds to the pointing error) and FWHM beam size of the 843 MHz source.
{Center}: false color image of {\it Spitzer}/IRAC (8, 4.5, and 3.6 $\mathrm{{\mu}m}$ for red, green, and blue, respectively) is shown with contours of {\it Herschel}/SPIRE 250 $\mathrm{{\mu}m}$ (red, 0.1, 0.15, and 0.2 $\mathrm{Jy~beam^{-1}}$) and {\it Spitzer}/MIPS~24 $\mathrm{{\mu}m}$ (blue, 0.5, 0.7, and 0.9 $\mathrm{MJy~sr^{-1}}$).
{Right}: $K_S$-band image by SIRIUS/IRSF is shown with contours of {\it Spitzer}/IRAC 3.6 $\mathrm{{\mu}m}$ (red, 0.2, 0.4, 0.6, and 0.8 $\mathrm{MJy~sr^{-1}}$). The blue cross and circle show the position (cross length corresponds to the astrometric error) and seeing size of the $U$-, $B$-, and $V$-band source.}
\label{figure:image}
\end{figure}

The other 1.1 mm source, MMJ0107, located at $\mathrm{\alpha}(J2000)=01^\mathrm{h}07^\mathrm{m}03^\mathrm{s}_{\cdot}7$ and $\mathrm{\delta}(J2000)=-73^\circ02'02''$ has a 1.1 mm flux density of $43.3{\pm}8.4$~mJy.
Possible systematic bias imposed by confusion noise is estimated by a simulation in which we place a 45 mJy point source in the signal map and extract it, and we repeat this 30,000 times. 
We find the mean extracted flux of 45.0$\pm$6.7~mJy, suggesting that the flux boosting caused by confusion noises is negligible at this flux level.
We found no clear counterpart in SIMBAD.
We use the multi-wavelength data toward MMJ0107, and find counterparts in the radio to near-IR. 
Figure \ref{figure:image} shows the multi-wavelength images at the position of MMJ0107.
The flux densities are summarized in Table \ref{table:flux}.
Among the radio bands we identified a counterpart only at 843~MHz.
We also find submm and IR counterparts in the SPIRE (250, 350, 500 $\mathrm{{\mu}m}$), MIPS~24 $\mathrm{{\mu}m}$ and the four IRAC bands. 
The {\it Spitzer}/IRAC source is located at $\mathrm{\alpha}(J2000)=01^\mathrm{h}07^\mathrm{m}02^\mathrm{s}_{\cdot}45$ and $\mathrm{\delta}(J2000)=-73^\circ01'59''_{\cdot}6$.
A near-IR ($J$, $H$, $K_S$) source at $\mathrm{\alpha}(J2000)=01^\mathrm{h}07^\mathrm{m}02^\mathrm{s}_{\cdot}40$ and $\mathrm{\delta}(J2000)=-73^\circ01'59''_{\cdot}5$ is closely associated with the {\it Spitzer} source.
The IRAC and near-IR objects have astrometric errors including the statistical and systematic errors of  0.3  and 0.2~arcsec respectively, therefore the centroid positions of the IRAC and near-IR objects are in agreement within the positional uncertainties.
On the other hand, an optical ($U$, $B$, and $V$) source at $\mathrm{\alpha}(J2000)=01^\mathrm{h}07^\mathrm{m}02^\mathrm{s}_{\cdot}08$ and $\mathrm{\delta}(J2000)=-73^\circ01'59''_{\cdot}3$ is located $1.7''$ and $1.4''$ away from the IRAC and near-IR sources.
Given the $UBV$ astrometry of $0.7''$, the positional offsets between the $UBV$ object and near-IR and mid-IR objects are significant, so the optical object is a possible lensing galaxy, because a lensing object is typically located within a few arcsec \citep{1994ApJ...435...49R, 2010Sci...330..800N} for a galaxy--galaxy lens system.

The 1.1 mm and submm emissions may arise from cold ($\sim$20 K) dust in the molecular cores residing in the SMC or the Galaxy.
If MMJ0107 were located at the SMC, the spatial extent of the 1.1 mm source ($\leq 35''$) would require the source size of $<$10~pc, which can be a compact molecular cloud in the SMC.
We estimate the dust temperature and gas mass assuming a molecular cloud in the SMC using the 1.1 mm to 250 $\mathrm{{\mu}m}$ data.
We use dust emissivity $\kappa_{250 \mathrm{{\mu}m}}=5~\mathrm{cm^2~g^{-1}}$ \citep{2001ApJ...554..778L}, index of emissivity ${\beta}=1.5$, and a gas-to-dust ratio of 700 \citep{2007ApJ...658.1027L}.
As a result of single component $\chi^2$-fitting, we get a dust temperature of $9.9\pm0.3$ K and gas mass of $(6.5{\pm}0.8)\times 10^5 {M_\sun}$.
However, it is unlikely that the 10 K dust component dominates in the scale of a molecular cloud \citep{2010A&A...524A..52B}.
Additionally, H{\small I} distribution \citep{1999MNRAS.302..417S} around MMJ0107 has no local peaks at the velocity of the SMC, implying that no molecular cloud exists.
Therefore, MMJ0107 is not likely to be located at the distance of the SMC.
The dust temperature of 10 K might also be explained by a molecular cloud core in our Galaxy.
In the SMC Wing region, however, there is no strong concentration of Galactic {\small I} gas, which is unnatural for a Galactic source.

We estimate the photometric redshifts by spectral energy distribution (SED) fitting of template galaxies.
We use four SED templates of galaxies: nearby starburst galaxies Arp~220 and M82 \citep{1998ApJ...509..103S}, averaged SED of 76 SMGs \citep{2010A&A...514A..67M}, and the well-known lensed galaxy SMM~J2135$-$0102 \citep{2010Natur.464..733S}.
We use the multi-wavelength data set at wavelengths longer than 20 $\mathrm{{\mu}m}$: MOST, ATCA (for upper limit), AzTEC/ASTE, {\it Herschel}/SPIRE, and {\it Spitzer}/MIPS (70/160 $\mathrm{{\mu}m}$ for upper limit) and {\it WISE} W4 for SED template fittings.
We do not use the optical data due to the significant offset from near-IR and mid-IR objects implying the lensing galaxy.
We also do not use the near-IR and mid-IR photometric data because these bands strongly depend on stellar synthesis models and dust extinction, and may possibly be contaminated by the foreground lensing object.

\begin{figure}
\epsscale{1.00}
\plotone{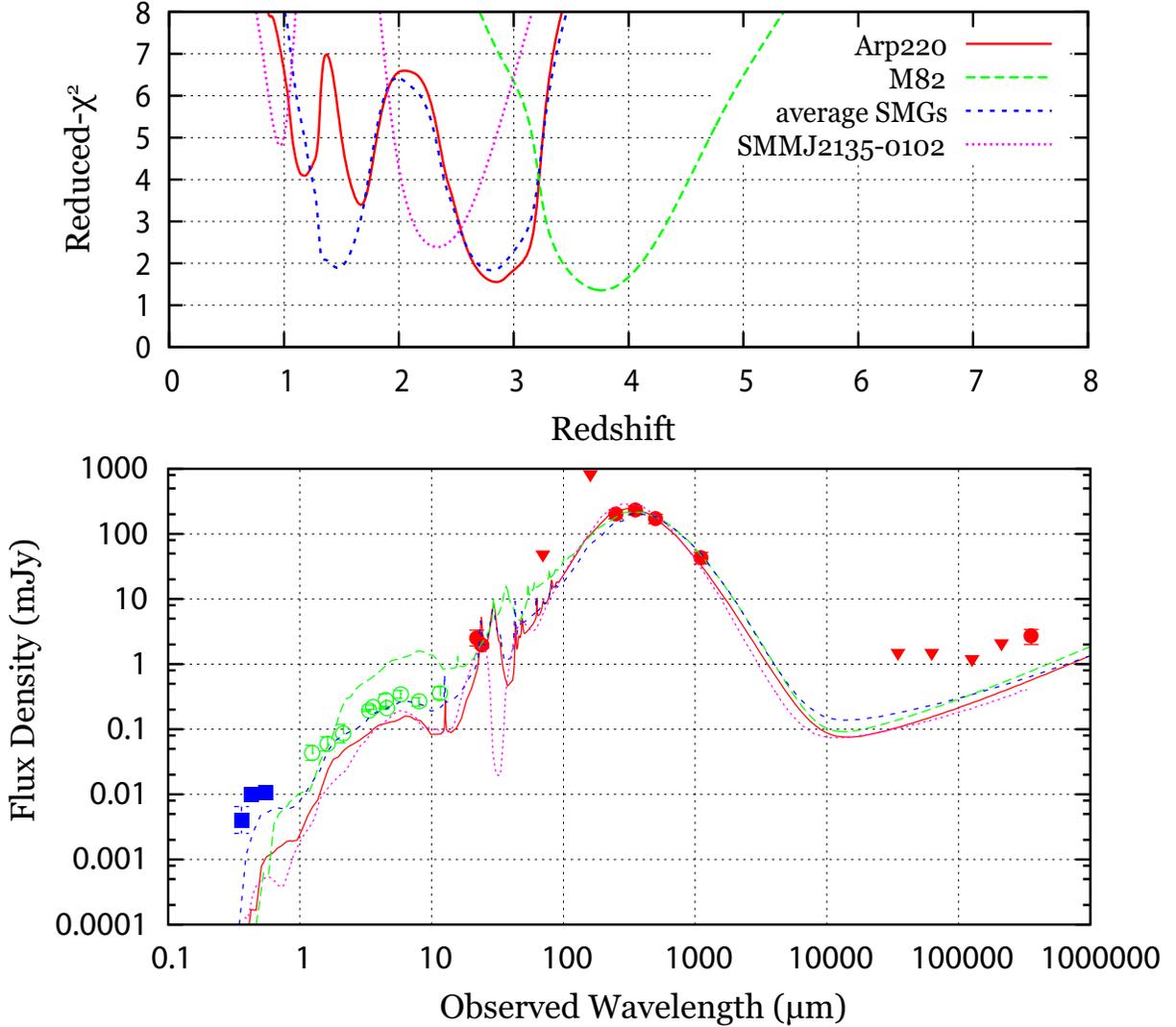}
\caption{
Result of SED fitting. {Top}: the reduced-$\chi^2$ value of each template SED fittings with redshift. {Bottom}: the photometric data points of MMJ0107 and template SEDs that minimize the $\chi^2$ values. 
Filled circles and down-pointed triangles are the flux densities and upper limits fit with the SED templates of SMGs, respectively.
Open circles are the flux densities that are not used for fitting with the SED templates.
Squares are the flux densities of the candidate lensing galaxy, which has significant positional offsets from the near-IR and mid-IR sources.
}
\label{figure:SED}
\end{figure}

Figure \ref{figure:SED} shows photometric redshift estimates.
The estimate using the M82 template gives a photometric redshift of $z=3.77^{+0.13}_{-0.14}$ and a FIR luminosity of $L_\mathrm{FIR}=1.9^{+0.3}_{-0.3}\times 10^{14}~{L_\sun}$, which shows the largest redshift and luminosity, and the least reduced-$\chi^2$ value among all templates.
The errors indicate the 68.3\% confidence intervals for statistical errors without consideration of systematic bias by the template selection.
The result of Arp~220 fitting shows $z=2.84{\pm}0.09$ and $L_\mathrm{FIR}=1.3^{+0.1}_{-0.2}\times 10^{14}~{L_\sun}$.
The averaged-SMGs template shows $z=2.80{\pm}0.09$ and $L_\mathrm{FIR}=1.2^{+0.2}_{-0.2}\times 10^{14}~{L_\sun}$, although the SED at $z=1.47\pm0.04$ and $L_\mathrm{FIR}=3.4^{+0.0}_{-0.2}\times 10^{13}~{L_\sun}$ is also probable.
The best-fit SED of the averaged SMGs is in good agreement with near-IR and mid-IR bands.
The SMM~J2135$-$0102 template shows $z=2.34^{+0.09}_{-0.10}$ and $L_\mathrm{FIR}=4.4^{+0.4}_{-0.5}\times 10^{13}~{L_\sun}$.
Thus, the photometric redshift estimated using the four SED templates ranges 1.4--3.9 and FIR luminosity of $(0.3$--$2.2) \times 10^{14}~{L_\sun}$.
The excesses of the 843 MHz photometry to the best-fit SED models are quantified by $q$-value, which is defined as $q=\log_{10}(L_\mathrm{IR}[{L_\sun}]/3.75\times 10^{12}/L_{\nu,~\mathrm{1.4~GHz}}[L_\sun~\mathrm{Hz^{-1}}])$ \citep{2010A&A...514A..67M}.
The $q$-values are estimated to be $2.07\pm^{0.13}_{0.10}$ for M82, $2.16\pm^{0.13}_{0.10}$ for Arp~220, $2.16\pm^{0.13}_{0.10}$ and $2.24\pm^{0.13}_{0.10}$ for the averaged-SMGs ($z=2.80$ and $z=1.47$, respectively), and $1.89\pm^{0.13}_{0.10}$ for SMM~J2135$-$0102, respectively.
The $q$-values of MMJ0107 are consistent with the dispersion of typical SMG samples, but relatively lower than the mean value \citep[$2.32{\pm}0.04$, dispersion $0.34$;][]{2010A&A...514A..67M}.
This may imply the existence of a radio-loud AGN, although this can also be explained by the photometric error of the 843 MHz data by confusion of unresolved radio sources, or enhancement of synchrotron emission expected for a spatially extended starburst \citep[e.g,][]{2010ApJ...717..196L}.

Here, we investigate the SFR of MMJ0107 from the FIR luminosity, following:
$$\mathrm{SFR} (M_\sun~\mathrm{yr^{-1}}) = 1.73 \times 10^{-10} L_\mathrm{FIR} ({L_\sun}),$$
which is a scaling law determined from local starburst galaxies \citep{1998ARA&A..36..189K}.
The inferred range in FIR luminosity corresponds to an SFR of $5600$--$39,000 M_\sun~\mathrm{yr^{-1}}$.
This extremely high SFR is unphysical, and can be accounted for by strong lensing magnification of a galaxy--galaxy lens system, although the SFR can be overestimated due to dust emission powered by an AGN.
If we assume the maximum magnification factor $\mu{\sim}40$, which has been reported by the Sloan Lens ACS Survey \citep{2011ApJ...734..104N}, this reduces the intrinsic SFR of $\lesssim~140$--$1000~M_\sun~\mathrm{yr^{-1}}$, which is similar to those often found in unlensed SMGs and ULIRGs.
Accordingly, an active star-forming galaxy highly magnified by a gravitational lens is the most likely explanation for MMJ0107.

\section{Discussion}
We found one ultra-bright SMG in the 1.21 $\deg^2$ field, corresponding to a number density of $0.83{\pm}0.83~\deg^{-2}$ ($S_{1.1\mathrm{mm}}>$ 43~mJy).
In addition, 1.1 mm deep surveys by the AzTEC instrument toward GOODS-N \citep{2008MNRAS.391.1227P}, SSA22 \citep{2009Natur.459...61T}, Lockman Hole \citep{2010MNRAS.401..160A}, GOODS-S \citep{2010MNRAS.405.2260S}, ADF-S \citep{2011MNRAS.411..102H}, SXDF \citep{2011MNRAS.411..102H}, and COSMOS \citep{2011MNRAS.415.3831A} found only one ultra-bright SMG \citep[Orochi, $37.3\pm0.7$~mJy;][]{2011MNRAS.415.3081I} in a total area of 2.0 $\deg^2$.
Thus, the cumulative number counts at $S_{1.1\mathrm{mm}}>$ 35~mJy are approximately $0.63{\pm}0.44~\deg^{-2}$ (i.e., two sources in a 3.2 $\deg^2$ region).
Here we compare this to the number count of ultra-bright galaxies detected by {\it Herschel}-ATLAS SDP sources.
Since the 500 $\mathrm{{\mu}m}$ to 1.1 mm flux ratio of MMJ0107 ($\sim$4) is smaller than those of {\it Herschel}-ATLAS SDP sources \citep[5--8;][]{2012ApJ...757..135L}, assuming the flux ratio of 4 as the lower limit, the flux density of 35~mJy at 1.1~mm corresponds to 140~mJy at 500 $\mathrm{{\mu}m}$. 
There are four lensed SMGs that have flux densities of $>$ 140~mJy at 500 $\mathrm{{\mu}m}$ in {\it{Herschel}}-ATLAS SDP field of 14.4 $\deg^2$ \citep{2010Sci...330..800N}, corresponding to a cumulative count of $N(S_{500\mathrm{{\mu}m}}>140 \mathrm{mJy}) = 0.28{\pm}0.14~\deg^{-2}$.
We also compare the 1.1 mm number density with the 1.4 mm one of SPT \citep{2010ApJ...719..763V}. \citet{2012MNRAS.423..575S} estimated the 1.1 mm counts using 1.4 mm data assuming a spectral index of $\alpha=2.65$. The flux density of $>$ 35~mJy at 1.1~mm corresponds to 19~mJy at 1.4~mm.
The number density of dust-dominated sources is $N(S_{1.4\mathrm{mm}}>17\mathrm{mJy})=0.14{\pm}0.05 \deg^{-2}$.
Accordingly, the number density of 1.1 mm selected ultra-bright SMGs is consistent with 500 $\mathrm{{\mu}m}$ and 1.4 mm counts.

We pointed out that the $UBV$ object located at $1.4''$ away from near-IR counterparts can be the lensing object.
In this case, the decrease in the flux at $U$ band is caused by the 4000\AA break, and the redshift of the lensing object can be $\lesssim~0.1$.
On the other hand, it is also possible that the near-IR emission arises from the lensing galaxy, although it is hard to estimate the photometric redshift because of the lack of a spectral feature in observed near-IR bands.
Alternatively, we estimate the redshift of the lensing object using a well-known correlation between $K$-band magnitude and redshift ($K$--$z$ relation) for the radio-selected massive elliptical galaxies \citep[][the $K$-band luminosity between 0.3 and 14 $L_\star$]{2003MNRAS.339..173W}.
The $K$-band luminosity range is reasonable for lensing galaxies: for example, strong lensing systems {\it Herschel}-ATLAS SDP~9, 17, and 81 \citep{2010Sci...330..800N} have the observed-frame $K$-band luminosities of $(3.5{\pm}0.8)\times 10^{11}$, $(6.4{\pm}1.4)\times 10^{11}$, and $(2.43{\pm}0.03)\times 10^{11}~L_\sun$ including the $k$-correction \citep{2001MNRAS.326..745M}, respectively.
The luminosities correspond to 2--6~$L_\star$ for the local luminosity function \citep[$1 L_\star=(1.1{\pm}0.1)\times 10^{11}~L_\sun$;][] {2001ApJ...560..566K}, respectively.
If we assume that the lensing galaxy of MMJ0107 is similar to ellipticals observed by \citet{2003MNRAS.339..173W}, then $K_S=17.23{\pm}0.25$ observed for MMJ0107 gives a redshift of $0.9{\pm}0.2$.
Therefore, the near-IR object can also be the lensing galaxy.
In this case, the $UBV$ object would be unrelated with this lens system, or even might be an interacting galaxy associated with MMJ0107 at $z \sim 3$, if the decrement at $U$ band is attributed to the Lyman break.
However, due to the lack of short wavelength data, it is difficult to further discuss the lensing object and the intrinsic properties of MMJ0107.
Additional near-IR and optical observations are needed.

\section{Conclusion and Future Prospects}
In this Letter, we have reported a new ultra-bright SMG behind the SMC.
The 1.1 mm flux density of MMJ0107 is 43.3$\pm$8.4~mJy, making it one of the brightest SMGs found in AzTEC/ASTE 1.1-mm continuum observations.
The SED fittings of MMJ0107 show the redshift of 1.4--3.9 and the FIR luminosity of $(0.3$--$2.2)\times{10^{14}}~\mu^{-1}~L_\sun$, which corresponds to the SFR of $5600$--$39000~\mu^{-1}M_{\sun}\mathrm{yr^{-1}}$, suggesting that MMJ0107 is a highly magnified lensed SMG.
Further spectroscopic observations of mm/submm emission lines for MMJ0107 will allow us not only to determine the redshift, but also to estimate the mass and physical state of the cool gas components, which makes it possible to constrain the star formation activities in MMJ0107.

This study demonstrates that a high-redshift galaxy can be reasonably separated from foreground objects through multi-band photometry, as described in Section \ref{l:result}. Further investigation in existing mm/submm data taken toward nearby galaxies and Galactic star-forming regions allows us to find more ultra-bright SMGs, which will improve the statistics of the brightest part of the number counts.

\acknowledgments
This work was partly supported by MEXT KAKENHI grant number 20001003 and JSPS KAKENHI grant numbers 19403005 and 23840007. 
K.S.S. is supported by NRAO, which is a facility of NSF operated under cooperative agreement by AUI.
The ASTE project is driven by NRO/NAOJ, in collaboration with the University of Chile, and Japanese institutes including the University of Tokyo, Nagoya University, Osaka Prefecture University, Ibaraki University, Hokkaido University, and Joetsu University of Education.
Observations with ASTE were carried out remotely from Japan using NTT's GEMnet2 and its partner R\&E networks, which are based on AccessNova collaboration between the University of Chile, NTT Laboratories, and NAOJ.
This research has made use of SIMBAD, operated at CDS, Strasbourg, France.
This work is based on data products made with the {Spitzer Space Telescope} (JPL/Caltech under a contract with NASA) and {WISE} (UCLA and JPL/Caltech funded by NASA).
SIRIUS/IRSF data provided by JVO (\url{http://jvo.nao.ac.jp/}) made it possible to complete this work.

{\it Facilities:} \facility{ASTE/AzTEC}, \facility{{\it Herschel}/SPIRE},  \facility{{\it Spitzer}/MIPS and IRAC}, \facility{{\it WISE}}, \facility{ATCA}, \facility{Molonglo Observatory}, \facility{IRSF/SIRIUS}, \facility{Swope/Great Circle Camera}.




\clearpage

\begin{table}
\begin{center}
\caption{Flux Densities of MMJ0107}
\begin{tabular}{lcccc}
\tableline\tableline
Instruments & Band & Flux Density & Beam Size& Reference \\
\tableline
\multicolumn{5}{l}{(Ultra-bright SMG)} \\										
MOST	&	843~MHz	&	2.7$\pm$0.7~mJy	&	45$''$	&	1	\\	
ATCA	&	1.42~GHz	&	$<$ 2.1~mJy\tablenotemark{a}	&	20$''$	&	1	\\	
	&	2.37~GHz	&	$<$ 1.2~mJy\tablenotemark{a} 	&	30$''$	&	1	\\	
	&	4.8~GHz	&	$<$1.5~mJy\tablenotemark{a}	&	35$''$	&	2	\\	
	&	8.64~GHz	&	$<$1.5~mJy\tablenotemark{a}		&	22$''$	&	2	\\	
AzTEC/ASTE	&	1.1~mm	&	43.3$\pm$8.4~mJy	&	35$''$	&	3	\\	
SPIRE/{\it Herschel}	&	500 $\mathrm{{\mu}m}$	&	172$\pm$28~mJy	&	36.3$''$	&	4	\\	
	&	350 $\mathrm{{\mu}m}$	&	233$\pm$37~mJy	&	24.9$''$	&	4	\\	
	&	250 $\mathrm{{\mu}m}$	&	203$\pm$32~mJy	&	18.2$''$	&	4	\\	
MIPS/{\it Spitzer}	&	160 $\mathrm{{\mu}m}$	&	$<$ 830~mJy\tablenotemark{a}	&	40$''$	&	5	\\	
	&	70 $\mathrm{{\mu}m}$	&	$<$ 49~mJy\tablenotemark{a}	&	18$''$	&	5	\\	
	&	24 $\mathrm{{\mu}m}$	&	1938$\pm$170 $\mathrm{\mu}$Jy	&	6$''$	&	5	\\	
IRAC/{\it Spitzer}	&	8 $\mathrm{{\mu}m}$	&	265$\pm$40 $\mathrm{\mu}$Jy	&	2$''$	&	5	\\	
	&	5.8 $\mathrm{{\mu}m}$	&	348$\pm$47 $\mathrm{\mu}$Jy	&	1.9$''$	&	5	\\	
	&	4.5 $\mathrm{{\mu}m}$	&	277$\pm$69 $\mathrm{\mu}$Jy	&	1.7$''$	&	5	\\	
	&	3.6 $\mathrm{{\mu}m}$	&	219$\pm$17 $\mathrm{\mu}$Jy	&	1.7$''$	&	5	\\	
{\it WISE}	&	22.1 $\mathrm{{\mu}m}$ (W4)	&	2511$^{+802}_{-608}$ $\mathrm{\mu}$Jy	&	12.0$''$	&	6	\\	
	&	11.6 $\mathrm{{\mu}m}$ (W3)	&	360.3$^{+88}_{-70}$ $\mathrm{\mu}$Jy	&	6.5$''$	&	6	\\	
	&	4.6 $\mathrm{{\mu}m}$ (W2)	&	213.9$^{+12}_{-11}$ $\mathrm{\mu}$Jy	&	6.4$''$	&	6	\\	
	&	3.35 $\mathrm{{\mu}m}$ (W1)	&	194.6$^{+8.0}_{-7.7}$ $\mathrm{\mu}$Jy	&	6.1$''$	&	6	\\	
SIRIUS/IRSF	&	2.14 $\mathrm{{\mu}m}$ ($K_S$)	&	85.5$^{+32.5}_{-23.6}$ $\mathrm{\mu}$Jy	&	1.1$''$	&	7	\\	
	&	1.63 $\mathrm{{\mu}m}$ ($H$)	&	59.5$^{+16.8}_{-13.1}$ $\mathrm{\mu}$Jy	&	1.2$''$	&	7	\\	
	&	1.25 $\mathrm{{\mu}m}$ ($J$)	&	43.5$^{+13.3}_{-10.2}$ $\mathrm{\mu}$Jy	&	1.3$''$	&	7	\\	\tableline	 
\multicolumn{5}{l}{(Lens candidate)} \\										
GCC/LCST
	&	0.55 $\mathrm{{\mu}m}$ ($V$)	&	10.7$^{+2.2}_{-2.4}$ $\mathrm{\mu}$Jy	&	1.5$''$	&	8	\\	
	&	0.45 $\mathrm{{\mu}m}$ ($B$)	&	10.0$^{+2.3}_{-1.9}$ $\mathrm{\mu}$Jy	&	1.5$''$	&	8	\\	
	&	0.37 $\mathrm{{\mu}m}$ ($U$)	&	4.0$^{+2.5}_{-1.5}$ $\mathrm{\mu}$Jy	&	1.5$''$	&	8	\\	
\tableline
\label{table:flux}
\end{tabular}
\tablenotetext{a}{3$\sigma$ upper limit.}
\tablerefs{
(1) \citet{2011SerAJ.183..103W};
(2) \citet{2010AJ....140.1511D};
(3) this paper;
(4) {\it Herschel} archive data;
(5) \citet{2011AJ....142..102G};
(6) \citet{2010AJ....140.1868W};
(7) \citet{2007PASJ...59..615K};
(8) \citet{2002AJ....123..855Z}.}
\end{center}
\end{table}
\clearpage

\end{document}